\documentstyle[12pt,epsfig]{article}
\thicklines

\title{Radiative leptonic $B$ decays in the instantaneous Bethe-Salpeter
approach}
\author{ G.~A.~Chelkov$^1$, M.~I.~Gostkin$^1$ and Z.~K.~Silagadze$^2$ 
\vspace*{3mm} \\
{\em $^1$ JINR, Laboratory of Nuclear Problems,} \\
{\em 141 980, Dubna, Russia;} 
\vspace*{2mm} \\
{\em $^2$ Budker Institute of Nuclear Physics,} \\
{\em 630 090, Novosibirsk, Russia;}
}

\date{}

\begin{document}
\large
\maketitle

\begin{abstract}
Rare radiative leptonic decay $B\to l\bar \nu_l \gamma$ is studied in the
instantaneous Bethe-Salpeter approach. The results are compared to other
relativistic quark model predictions.
\end{abstract}

The $B$-meson decay constant $f_B$ is an important phenomenological
parameter, not easy to measure directly. Purely leptonic decays
$B\to l\bar \nu_l$, from which it could be extracted in principle, suffer
either from helicity suppression $m_l^2/M_B^2$ for light leptons, or
from reconstruction difficulties for $\tau$-channel owing to the presence
of two neutrinos in the final state. Decay rates expected in the Standard 
Model are
\begin{eqnarray}
\Gamma(B\to l\bar \nu_l)=\frac{G_F^2}{8\pi}|V_{ub}|^2f_B^2
\frac{m_l^2}{M_B^2}  M_B^3 \left ( 1-\frac{m_l^2}{M_B^2}\right )^2
\approx \left \{ \matrix {
7\cdot 10^{-12}, \; {\rm if} \; l=e^- \cr
3\cdot 10^{-7}, \; {\rm if} \; l=\mu^- } \right .
\nonumber \end{eqnarray}
\noindent where numbers quoted correspond to $|V_{ub}|=3\cdot 10^{-3}, \;
f_B=200 MeV $ and $ \tau_B\approx 1.65 ps$. The present experimental limits 
\cite{1} on these decay rates are at least an order of magnitude larger.

Some times ago Burdman, Goldman and Wyler (BGW) suggested an alternative,
although model-dependent way for $f_B$ measurement \cite{2}. The crucial
observation was that the helicity suppression can be overcome and turned
into an $\alpha$ (the e.m. fine-structure constant) suppression by an 
additional photon emission in radiative weak decays 
$B\to l\bar \nu_l \gamma$. The dominant contribution in these radiative 
decays comes from the $B^*$-pole intermediate state: a spin-0 $B$ meson 
emits a hard photon and transforms into an off-shell spin-1 $B^*$ meson 
which by itself undergoes weak decay without the helicity suppression. 
Thus measurement of these decay rates gives a tool to access $B^*$ meson 
decay constant $f_B^*$. Heavy quark symmetry can be used then to relate 
$f_B^*$ and $f_B$.

Afterwards $B\to l\bar \nu_l \gamma$ decay ( up to $m_l^2/M_B^2$ accuracy
the decay rate is independent of the lepton flavor) was considered in a
number of publications. The BGW analysis was further refined by Colangelo,
De Fazio and Nardulli \cite{3}. It was found, in particular, that the
axial-vector ($B^\prime$) intermediate state contributes about 10\% in the
total decay rate. Atwood, Eilam and Soni used a simple nonrelativistic quark
model to estimate the decay width and obtained \cite{4}
$Br(B\to l\bar \nu_l \gamma)\approx 3.5\cdot 10^{-6}$, about 12 times larger
than the purely leptonic branching ratio $Br(B\to \mu\bar \nu_\mu )$.
Subsequent relativistic generalizations involve the use of: spinless
Salpeter equation \cite{5}, light front dynamics \cite{6}, light cone QCD
sum rules \cite{7}, perturbative QCD combined with the heavy quark effective
theory \cite{QCD}. All of them confirm the main conclusion about the
radiative decay mode enhancement, although with smaller branching ratio
ranging from $0.9\cdot 10^{-6}$ \cite{5} to $2\cdot 10^{-6}$ \cite{7}. The
recent experimental upper limits \cite{8} are still far above of these
predictions. 

\begin{figure}[htb]
\resizebox{0.99\textwidth}{!}{
\includegraphics{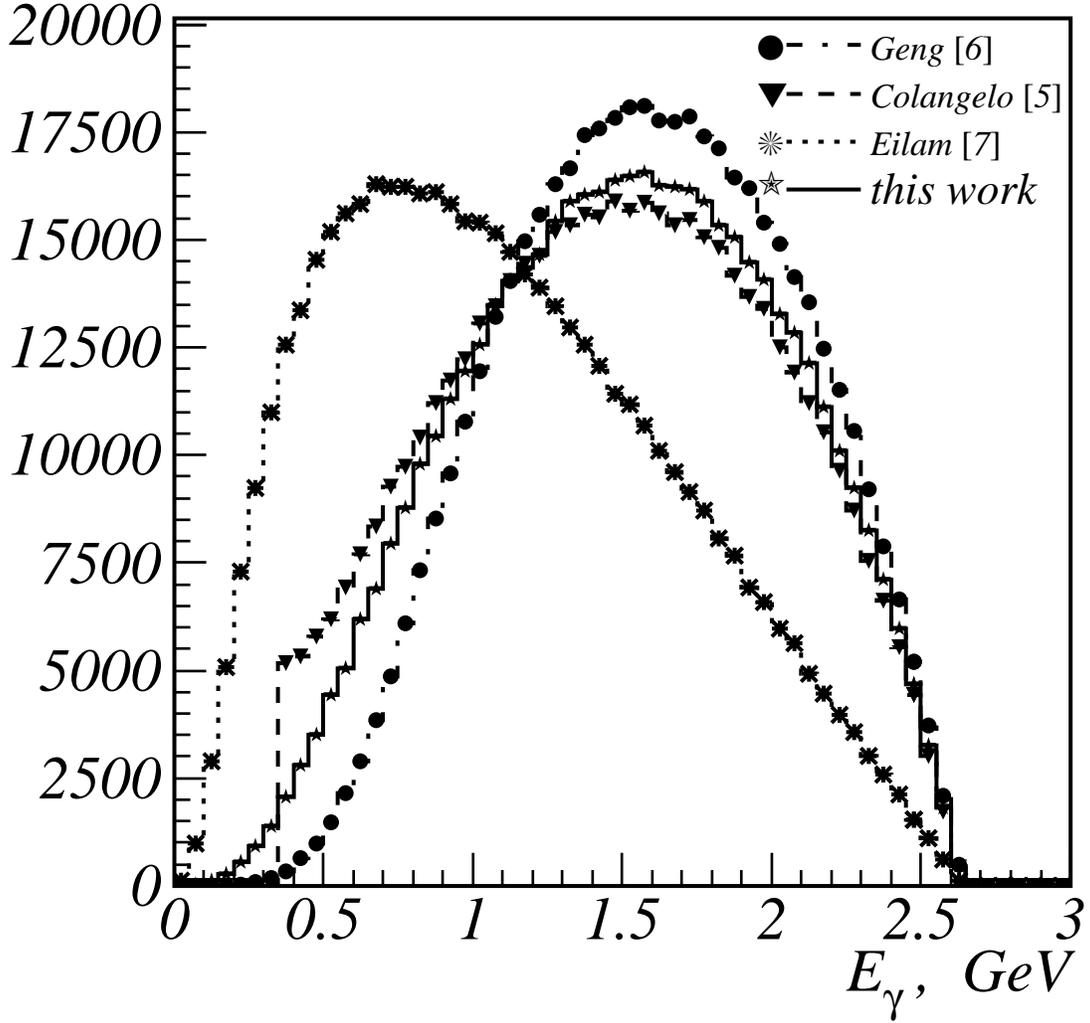}
}
\caption{Photon spectra in various relativistic quark models for 
the $B\to l\bar \nu_l \gamma$ decay.}
\label{fig1}
\end{figure}

It is worthwhile to note that the photon spectra in various
relativistic quark models are significantly different as illustrated by
Fig.\ref{fig1}.
An inverted-parabolic shape from \cite{4}, with a mean value
$\sim 1.3 GeV$, is asymmetrically
modified in \cite{5,6} towards higher photon energies, but in opposite
direction in the light cone QCD sum rules approach
\cite{7}, with the mean value shifted down to $\sim 0.8 GeV$. 
This softening of the
photon spectrum can effect the signal detection efficiency \cite{8,BaBar} 
and the signal-background separation. We decided to check whether  
character of change in
the photon energy distribution, predicted in \cite{7}, still persists in an
another, instantaneous Bethe-Salpeter approach based, relativistic quark
model \cite{9,10,11}, which was successfully applied earlier to describe
light \cite{12} and heavy \cite{11,13} meson spectra, as well as
electromagnetic form-factors of the ground state pseudoscalar and vector
mesons \cite{14}, various two photon widths \cite{15} and heavy meson weak
decays \cite{16}. This approach respects the heavy quark spin symmetry 
in the limit  $m_b \to \infty$ \cite {11}. It also incorporates the 
spontaneous breaking of chiral symmetry in the light flavor sector, which 
happens due to the generation of a dynamical quark mass in the light quark
self--energy generated by the interaction potential.

The $B\to l\bar \nu_l \gamma$ matrix element has the following general
structure
\begin{eqnarray}
A(B\to l\bar \nu_l \gamma)=\frac{ieG_FV_{ub}}{\sqrt{2}M_B}
\epsilon^*_\nu H^{\mu\nu}~\bar u(p_l)\gamma_\mu(1-\gamma_5)v(p_\nu) \; ,
\nonumber \end{eqnarray}
\noindent where $H_{\mu\nu}$, the hadronic tensor, is
uniquely determined, due to gauge invariance and Lorentz covariance, by
two invariant form-factors $F_V$ and $F_A$:
\begin{eqnarray}
H_{\mu \nu}=F_A~[g_{\mu\nu}P\cdot k-k_\mu P_\nu ]+
iF_V~\epsilon_{\mu\nu\sigma\tau}P^\sigma k^\tau 
\label{eq1} \end{eqnarray} 
\noindent $P$ being the $B$-meson 4-momentum and $k$ -- the photon
4-momentum (in the following we will assume the $B$-meson rest frame in all
expressions, so $P=(M_B,\vec{0})$~).

The standard procedure leads to the following differential width ($m_l$ is
neglected)
$$ \frac{d^2\Gamma}{dx_1~dx_2}= \vspace*{-3mm} $$
\begin{eqnarray}
M_B\frac{\alpha(G_FM_B^2)^2}{16\pi^2}
|V_{ub}|^2~[\rho_+(x_1,x_2)|F_V+F_A|^2+\rho_-(x_1,x_2)|F_V-F_A|^2]  
\label{eq2} \end{eqnarray}
\noindent where $\rho_+(x_1,x_2)=(1-2x_1)(1-2x_2)^2, \;$
$\rho_-(x_1,x_2)=(1-2x_1)(1-2x_3)^2$ and
$x_1=E_\gamma/M_B,\; x_2=E_\nu/M_B,\;
x_3=E_l/M_B=1-x_1-x_2$ are the photon, neutrino and charged lepton
energy fractions. $F_V$ and $F_A$ form-factors depend only on the photon
energy, so one integration can be readily done while calculating the decay
width, and we get
$$ \Gamma(B\to l\bar \nu_l \gamma)= \vspace*{-3mm} $$
\begin{eqnarray} 
M_B\frac{\alpha(G_FM_B^2)^2}{6\pi^2}|V_{ub}|^2
\int\limits_0^\frac{1}{2}x_1^3(1-2x_1)~[~|F_V(x_1)|^2+|F_A(x_1)|^2~]~dx_1  
\; . \label{eq3} \end{eqnarray}

The photon emission from the initial light quark gives the most important
contribution to the $B\to l\bar \nu_l \gamma$ decay amplitude \cite{4}:

\begin{picture}(120,55)(-15,0)
\put(11,25){$b$}
\put(11,19){$\bar u$}
\put(15,25){\vector(2,0){6}}
\put(21,25){\line(2,0){7}}
\put(28,20){\vector(-2,0){10}}
\put(18,20){\line(-2,0){3}}
\put(28,22.5){\circle*{5}}
\put(28,25){\vector(4,3){8}}
\put(36,31){\line(4,3){8}}
\put(44,8){\vector(-4,3){8}}
\put(36,14){\line(-4,3){8}}
\put(44,37){\vector(0,-2){14.5}}
\put(44,22.5){\line(0,-2){14.5}}
\multiput(45.5,8)(6,0){6}{\oval(3,3)[t]}
\multiput(48.5,8)(6,0){6}{\oval(3,3)[b]}
\put(46.5,37){\oval(5,3)}
\put(49,37){\vector(3,1){15}}
\put(64,42){\line(3,1){15}}
\put(79,27){\vector(-3,1){15}}
\put(64,32){\line(-3,1){15}}
\put(28,30){P}
\put(32,33){+}
\put(36,36){q}
\put(32,11){q}
\put(46,22.5){k+q}
\put(81,7){$\gamma$}
\put(60,2){k,$\epsilon$}
\put(81,26){$\bar\nu_l$}
\put(81,46){l}
\end{picture}


From this diagram the corresponding contribution to the hadronic tensor is
easily obtained
\begin{eqnarray}
H_{\mu\nu}= 
\label{eq4} \end{eqnarray}
$$\vspace*{-3mm} -M_BN_cQ_u \int{ \frac{dq}{(2\pi)^4}Sp\left \{ \gamma_\mu
(1-\gamma_5)G_{(b)}(P+q)\Gamma(\vec{q};P)G_{(u)}(q)\gamma_\nu G_{(u)}(q+k)
\right \}} , \vspace*{5mm}$$ 
\noindent where $Q_u=\frac{2}{3}$, $G_{(q)}(p)=\frac{i}
{\hat p -m_q}$ stands for the constituent
$q$-quark propagator and $\Gamma(\vec{q};P)$ -- for the Bethe-Salpeter
vertex function \cite{10}. The Mandelstam formalism \cite{17} or the bilocal
effective meson theory \cite{18} gives a general guidelines how to calculate
transition or decay amplitudes, involving bound states, in terms of this
vertex function.

In the instantaneous approximation the Bethe-Salpeter vertex function
$\Gamma(\vec{q};P)$ depends only on the relative three-momentum $\vec{q}$
\cite{10}. So the dependence on $q_0$ in (4) is completely due to quark
propagators and the $q_0$-integration may be performed analytically using
the residue theorem and
$$G_{(q)}(p)=i\left [ \frac{\Lambda_+^{(q)}(\vec{p})}{p_0-\omega_{(q)}
(\vec{p})+i\epsilon}+\frac{\Lambda_-^{(q)}(\vec{p})}{p_0+\omega_{(q)}
(\vec{p})-i\epsilon} \right ] \gamma^0 \; , $$
\noindent where $\omega_{(q)}(\vec{p})=\sqrt{m_q^2+\vec{p}~^2}$ and
$\Lambda^{(q)}_\pm$ are the standard projection operators on positive and
negative energies
$$\Lambda^{(q)}_\pm (\vec{p})=\frac{1}{2}\left ( 1\pm 
\frac{\vec{\alpha}\cdot \vec{p}
+\beta m_q}{\omega_{(q)}(\vec{p})}\right ).$$

After $q_0$-integration in (4) we obtain several terms, from which the
leading contribution comes from the following one 
\begin{eqnarray}
H_{\mu\nu}\approx 
- \int{ \frac{d\vec{q}}{(2\pi)^3}~\frac {M_BN_cQ_u~
Sp\left \{ \gamma_\mu
(1-\gamma_5)\Gamma_{+-}(\vec{q};P)\gamma_\nu \Lambda_+^{(u)}(\vec{q}+\vec{k})
\gamma_0 \right \} } {[M_B-\omega_{(b)}(\vec{q})-\omega_{(u)}(\vec{q})]
[k_0-\omega_{(u)}(\vec{q})-\omega_{(u)}(\vec{q}+\vec{k})] } } \; ,
\label{eq5} \end{eqnarray}
\noindent because in the heavy $b$-quark limit the characteristic momentum
scale for $\Gamma(\vec{q};P)$ is much less than $m_b$ and so we may use
$$ 
M_B\approx m_b \approx \omega_{(b)}(\vec{q}) \; , \; \;
\omega_{(u)}(\vec{q}+\vec{k}) \approx k_0 \gg \omega_{(u)}(\vec{q}) \; . 
$$

In (5) $\Gamma_{+-}(\vec{q};P)=\Lambda_+^{(b)}(\vec{q})\gamma^0\Gamma
(\vec{q};P)\gamma^0\Lambda_-^{(u)}(-\vec{q})$ and it can be immediately
replaced by the Salpeter wave function $\Phi(\vec{q})$ according to the
Salpeter equation \cite{10,11}
$$\Gamma_{+-}(\vec{q};P)=-[M_B-\omega_{(b)}(\vec{q})-\omega_{(u)}(\vec{q})]
\Lambda_+^{(b)}(\vec{q})\Phi(\vec{q})\Lambda_-^{(u)}(-\vec{q}).$$
\noindent In the heavy quark limit $\Phi(\vec{q})$ has the following simple
form \cite{11}
\begin{eqnarray}
\Phi(\vec{q})\approx \frac{l(q)}{q\sqrt{4\pi}}(1+\gamma_0)\gamma_5
S^{-1}_{(u)}(\vec{q}) \; , q=|\vec{q}| \; ,
\label{eq6} \end{eqnarray}
\noindent where $l(q)$ obeys the radial Salpeter equation and
$S^{-1}_{(u)}(\vec{q})$ is the inverse of the Foldy-Wouthuysen matrix for
$u$-quark:
$$S^{-1}_{(u)}(\vec{q})=\cos{\nu_{(u)}(q)}-\frac{\vec{q}\cdot\vec{\gamma}}
{|\vec{q}|}\sin{\nu_{(u)}(q)}$$
\noindent the Foldy-Wouthuysen angle being determined as follows
$$\cos{2\nu_{(u)}(q)}=\frac{m_u}{\omega_{(u)}(\vec{q})} \;, \; \; \;
\sin{2\nu_{(u)}(q)}=\frac{|\vec{q}|}{\omega_{(u)}(\vec{q})} \; . $$
\noindent To simplify (5), note also that
$$\Lambda_+^{(u)}(\vec{k}+\vec{q})\gamma_0=S_{(u)}^{-1}(\vec{k}+\vec{q})
\frac{1}{2}(1+\gamma_0)S_{(u)}(\vec{k}+\vec{q})\gamma_0=
\frac{1}{2}(S_{(u)}^{-2}(\vec{k}+\vec{q})+\gamma_0). $$
\noindent But
$$S_{(u)}^{-2}(\vec{k}+\vec{q})=\frac{m_u}{\omega_{(u)}(\vec{k}+\vec{q})}-
\frac{(\vec{k}+\vec{q})\cdot \vec{\gamma}}{\omega_{(u)}(\vec{k}+\vec{q})}
\approx -\frac{\vec{k}\cdot \vec{\gamma}}{\omega_{(u)}(\vec{k}+\vec{q})}
$$
\noindent and
$$S_{(u)}^{-2}(\vec{k}+\vec{q})+\gamma_0\approx
\frac{1}{\omega_{(u)}(\vec{k}+\vec{q})}
[\gamma_0~\omega_{(u)}(\vec{k}+\vec{q})-\vec{k}\cdot \vec{\gamma}]\approx
\frac{\hat k}{\omega_{(u)}(\vec{k}+\vec{q})} \; . $$

\noindent After these approximations (5) takes manifestly gauge invariant 
form:
\begin{eqnarray}
H_{\mu\nu}=-\frac{M_BN_cQ_u}{4\sqrt{\pi}}\int \frac{d\vec{q}}{(2\pi)^3}
~\frac{l(q)}{q}~\frac{Sp\{ \gamma_\mu (\gamma_5-1)(1-\gamma_0)S^
{-1}_{(u)}(\vec{q})
\gamma_\nu \hat{k} \} }{\omega_{(u)}(\vec{q})\omega_{(u)}(\vec{k}+\vec{q})}
\; . \label{eq7} \end{eqnarray}

Now it is straightforward to get $F_V$ and $F_A$, the invariant form-factors
from (7) as
\begin{eqnarray}
F_V=F_A=Q_u \left ( f(x_1)+g(x_1)\right )
\label{eq8}\end{eqnarray}
\noindent where (we have introduced $r=\frac{m_u}{M_B}$ and
$x=\frac{q}{M_B}$ dimensionless variables)
$$\hspace*{-10mm} f(x_1)=\frac{N_c}{(2\pi)^{5/2}}~\frac{1}{x_1}
\int\limits_0^\infty
l(x)[\varphi(x,x_1)-\varphi(x,-x_1)]\left [\frac{r+\sqrt{r^2+x^2}}
{\sqrt{r^2+x^2}} \right ]^{1/2}dx \vspace*{-3mm} $$
\begin{eqnarray}
g(x_1)=\frac{N_c}{(2\pi)^{5/2}}~\frac{1}{3x_1^2}\int\limits_0^\infty
l(x)[\psi(x,-x_1)-\psi(x,x_1)]\left [\frac{\sqrt{r^2+x^2}-r}
{\sqrt{r^2+x^2}} \right ]^{1/2}\frac{dx}{x}   ,
\label{eq9} \end{eqnarray}
\noindent and
\begin{eqnarray} 
\varphi(x,x_1)=\sqrt{\frac{r^2+(x+x_1)^2}{r^2+x^2}} \; , \; \;
\psi(x,x_1)=(r^2+x^2+x_1^2-xx_1)\varphi(x_1) \; .
\label{eq10} \end{eqnarray}

Note that in the nonrelativistic (heavy $u$-quark, but $m_u \ll m_b$)
limit characteristic momentum scale for $l(x)$ is much less than $r$. So 
$g(x_1) \to 0$ and
$$f(x_1) \to \frac{2\sqrt{2}N_c}{(2\pi)^{5/2}}~\frac{1}{x_1r}
\int\limits_0^\infty xl(x)dx \; .$$
\noindent In this limit $B$-meson decay constant $f_B$ takes the form
\cite{11}
$$f_B \approx \frac{4\sqrt{2}N_c}{(2\pi)^{5/2}}~M_B
\int\limits_0^\infty xl(x)dx \; .$$
\noindent So
$$f(x_1) \approx \frac{1}{2x_1}~\frac{f_B}{m_u} \; . $$
\noindent Substituting this into (8) and (3), we reproduce Atwood, Eilam and
Soni's result \cite{4}
$$ \Gamma(B\to l\bar \nu_l \gamma)=
M_B\frac{\alpha(G_FM_B^2)^2}{288\pi^2}Q_u^2|V_{ub}|^2\frac{f_B^2}{m_u^2}
\; . $$

The equality of the vector and axial current form factors, given by 
(\ref{eq8}), is surprising because in the pole approximation they receive
contributions from intermediate states with opposite parities. Nevertheless
this additional interesting spin symmetry was shown to hold in the high recoil
$E_\gamma\gg \Lambda_{QCD}$ region at least at one-loop order
by explicit leading twist perturbative QCD calculation \cite{QCD}.

The radial wave function $l(x)$ is determined from the integral equation which
in the heavy $b$-quark limit takes the form \cite{11}
$$ [M-m_b-\omega_{(u)}(p)]~l(p)= 
\frac{1}{2}\int\limits_0^\infty dq\left [
\sqrt{\left (1+\frac{m_u}{\omega_{(u)}(p)}\right )
\left (1+\frac{m_u}{\omega_{(u)}(q)}\right )}v_0(p,q)+ \right . $$
\begin{equation} \left .
\sqrt{\left (1-\frac{m_u}{\omega_{(u)}(p)}\right )
\left (1-\frac{m_u}{\omega_{(u)}(q)}\right )}v_1(p,q)\right ] l(q),   
\label{eq11} \end{equation} 
\noindent where $M$ is the bound state mass and $v_L(p,q)$ angular 
matrix element of the potential kernel is determined through 
$$\frac{pq}{(2\pi)^3}\int d\Omega_p \int d\Omega_q Y^*_{L^\prime M^\prime}
(\hat {\vec{p}})V(\vec{p}-\vec{q})Y_{L M}(\hat {\vec{q}})=v_L(p,q)\delta_
{L L^\prime}\delta_{M M^\prime}. $$ 
\noindent For estimation purposes we have used the same parameter set for
the $B$-meson description as given in \cite{11}. That is constituent quark 
masses $m_b=4.79~\mathrm{GeV},\; m_u = 0.33~\mathrm{GeV}$ and the linear
plus Coulomb  potential
$$V(r)=-\frac{4}{3}~\frac{\alpha_s}{r}+\sigma^2 r, $$
\noindent with $\alpha_s=0.39$ and $\sigma= 0.41~\mathrm{GeV}$.

\begin{figure}[htb]
\resizebox{0.95\textwidth}{!}{
\includegraphics{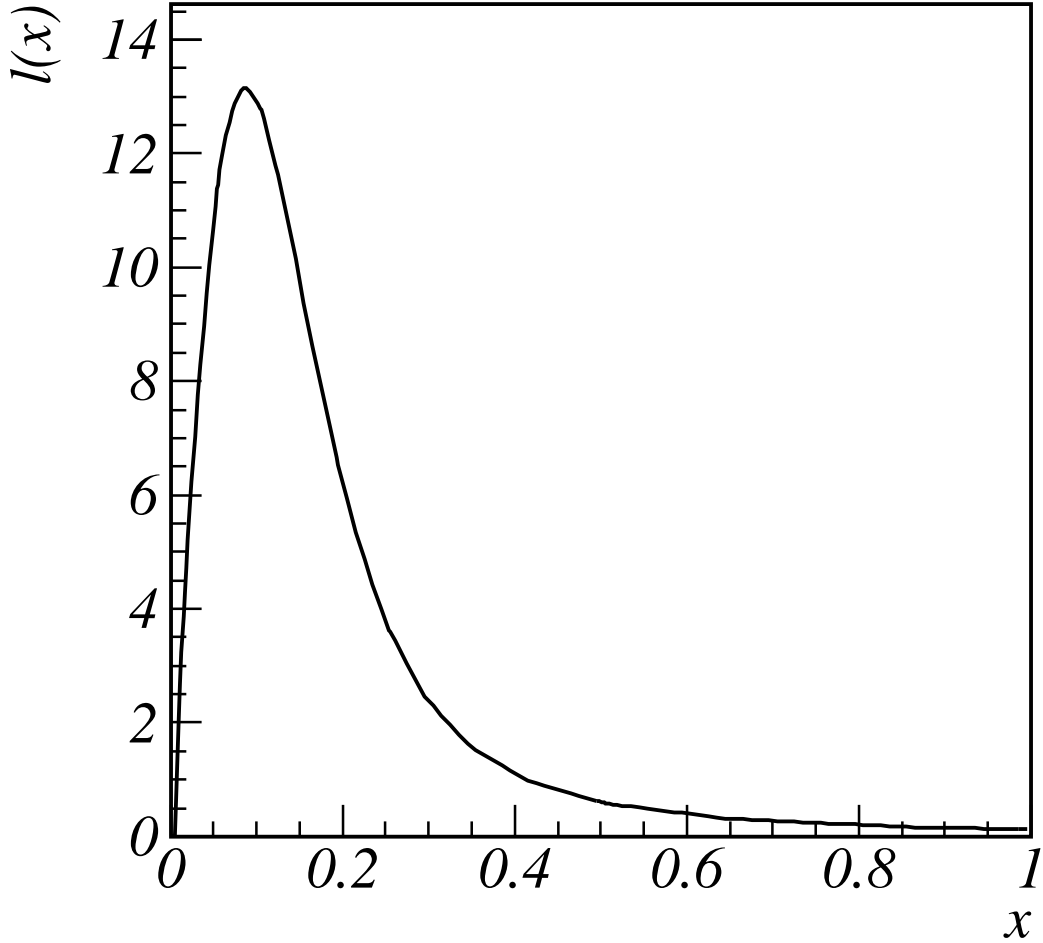}
}
\caption{The radial wave function $l(x)$.   
}
\label{fig2}
\end{figure}

For these parameters the radial Salpeter equation (\ref{eq11}) was solved
by Multhopp method \cite{11,19}. The resulting radial wave function is shown
in Fig.\ref{fig2}.
\noindent Note that (\ref{eq11}) determines $l(x)$ only up to normalization
constant. The normalization condition was considered for a general case in
\cite{11} and for our approximations looks like
$$\frac{N_c}{2\pi^3}\int\limits_0^\infty l^2(x) dx = 1.$$

Having at hand the radial wave function, we can calculate the decay branching
ratio
$$Br(B\to l\bar \nu_l \gamma)\approx 0.9\times 10^{-6}.$$
\noindent The resulting photon spectrum is indicated in Fig.\ref{fig1}. As
one can see, our results are very close to predictions of the spinless
Salpeter equation model \cite{5}, except in the low-energy part of the photon 
spectrum where the results of \cite{5} contain unphysical divergence and are
not reliable.

To summarize, the instantaneous Bethe-Salpeter model, considered in this 
article, gives a photon spectrum similar to other relativistic quark models
\cite{5} and \cite{6}, but different from the light cone QCD sum rules 
approach \cite{7}. The predicted branching ratio is within the reach of 
the BaBar experiment \cite{BaBar}. So we may expect that some experimental 
information will appear about this rare decay mode in the near future. To 
extract interesting quantities like $f_B$ from this information, a detailed 
understanding of the model uncertainties in simulation of this decay is 
necessary. We hope that our investigation will be useful in such studies.

\vskip 0.4cm
\noindent
{\bf Acknowledgement}
\vskip 0.4cm
\noindent
We thank Fulvia De Fazio and Chao-Qiang Geng for correspondence. The solution
of the radial Salpeter equation was provided by Christian Weiss. We gratefully
acknowledge this contribution and thank him for critical remarks.

\end{document}